\documentclass[final,3p]{elsarticle}

\usepackage{subcaption}
\usepackage{graphicx}
\usepackage{multirow}

\usepackage{pgfplots}
\usepackage{amsmath,amsthm,bm}
\usepackage{amssymb}
\usepackage[toc,page]{appendix}
\usepackage{chngcntr}
\usepackage{caption}
\usepackage{subcaption}
\usepackage{multirow}
\usepackage{verbatim}
\usepackage[]{algorithm2e}
\usepackage{float}
\usepackage{tabularx}
\usepackage{graphicx}
\usepackage{indentfirst}
\usepackage{tikz-3dplot}
\usepackage{etoolbox}

\usetikzlibrary{external}
\journal{Nuclear Inst. and Methods in Physics Research, A}

\tikzstyle{block} = [draw, rectangle, minimum height=3em, minimum width=6em]
\tikzstyle{sum} = [draw, circle, node distance=1cm]
\tikzstyle{input} = [coordinate]
\tikzstyle{output} = [coordinate]
\tikzstyle{pinstyle} = [pin edge={to-,thin,black}]
\newlength\figureheight
\newlength\figurewidth
\newcommand{\iu}{\text{i}}

\begin{document}

\begin{frontmatter}

\title{Local field reconstruction from rotating coil measurements in particle accelerator magnets}


\author[1,2]{Ion Gabriel Ion\corref{cor1}}
\ead{ion@temf.tu-datmstadt.de}
\author[3]{Melvin Liebsch}
\author[1,2,4]{Abele Simona}
\author[1,2]{Dimitrios Loukrezis}
\author[3]{Carlo Petrone}
\author[3]{Stephan Russenschuck}
\author[1,2]{Herbert De Gersem}
\author[1,2]{Sebastian Sch\"ops}
\cortext[cor1]{Corresponding author}

\address[1]{Institut für Teilchenbeschleunigung und Elektromagnetische Felder, Technische Universit\"at Darmstadt}
\address[2]{Graduate School Computational Engineering, Technische Universit\"at Darmstadt}
\address[3]{European Organization for Nuclear Research, CERN, Geneva, Switzerland}
\address[4]{MOX, Dipartimento di Matematica, Politecnico di Milano}

\begin{abstract}
In this paper a general approach to reconstruct three dimensional field solutions in particle accelerator magnets from distributed magnetic measurements is presented. To exploit the locality of the measurement operation a special discretization of the Laplace equation is used. Extracting the coefficients of the field representations yields an inverse problem which is solved by Bayesian inversion. This allows not only to pave the way for uncertainty quantification, but also to derive a suitable regularization. The approach is applied to rotating coil measurements and can be extended to any other measurement procedure.
\end{abstract}

\begin{keyword}
Accelerator Magnets\sep Magnetic Measurement \sep Bayesian inverse problems 
\end{keyword}

\end{frontmatter}

\section{Introduction}
Beam-physics simulations require evaluations of magnetic potentials or flux densities in the neighbourhood of a reference trajectory of the particle beam. Consequently, a suitable representation of the magnetic field has to be found. A local field description is needed whenever the effect of longitudinal field components is not neglectable for particle beam stability and dynamics calculations. Three dimensional field maps are required for spectrometer magnets, where the ratio between aperture and length does not admit a homogenity region inside the magnet \cite{Kazantseva_2019}. Recent studies also investigate the impact of fringe fields in focusing quadrupole magnets to nonlinear beam dynamics \cite{PUGNAT2020164350,SIMONA201933}. 
While the latter example is still in the phase of study, the longitudinal field profiles must be measured to determine the need and efficiency of correction strategies \cite{PUGNAT2020164350}. \par
For this reason, we present an approach to reconstruct the three-dimensional longitudinal field distribution in a cylindrical region inside the magnet's air gap, based on sampled measurement data. 
We make use of a rotating coil, short in length with respect to the magnets extend, at different longitudinal positions to infer magnetic field information and reconstruct a field solution in the air gap of the magnet satisfying Maxwell's equations. This extends the work in \cite{VENTURINI1999387}, but we complement the analysis with two important features: (i) a novel discretization of the Laplace problem \cite{simona2020}, to avoid coupling of local measurement errors with globally supported field solutions, and (ii) the methodological framework to propagate measurement uncertainties towards field quantities, by means of Bayesian inversion \cite{bayes_inv}. As this field representation is suited for fast evaluations and implies the regularity conditions of the magnetic field, it constitutes a direct link between particle beam dynamics and measurement data. This approach can be extended to other field representations, such as harmonic expansions or boundary element methods. 
The inverse problem, i.e., the extraction of the expansion coefficients from measurement data, is ill-posed and requires a suitable regularisation, which is derived by Bayesian inversion.\par
In section \ref{sec:coil_meas} we give an introduction on rotating coil measurements with focus on radial coil arrays. In section \ref{sec:field_rec_dPhi_k} we introduce the disctrization technique, based on spline basis functions along the magnets longitudinal axis. We then focus on the field solutions extraction from measurement data and uncertainty quantification in section \ref{sec:bayesian_inversion}. Finally, the theory is applied to measurement data taken in a dipole magnet in section \ref{sec:meas_corrector_dipole}.  


\section{Rotating Coil Measurements} \label{sec:coil_meas}
Rotating coils are among the most popular devices for field measurement in accelerator magnets. They can be easily calibrated in reference magnets and they determine the field harmonics with a high accuracy. 
Looking at multipole errors in the range of units of 10 000 with respect to the main field component, any imperfection in the coils rotational motion will introduce spurious harmonics in the measured signals. Unfortunately, small vibrations are inevitable in a real rotating coil setup. For this reason, special treatment to suppress the influences of transversal and torsional vibrations is indispensable for rotating coil measurements in the $<10^{-4}$ accuracy. As shown in \cite{Jain:1246517} in detail, one can combine the signals of multiple coils on the shaft in ways to reduce the sensors sensitivity towards specific multipole components. Applying such compensation for the main and lower multipole contribution yields compensation schemes for spurious solutions from transversal and torsional vibrations and also shaft deformations \cite{SORTI2020164599}. Of course, a perfect main field suppression would require identical geometrical properties of the coils involved. 
Printed-circuit-board (PCB) technology allows for highly accurate coil manufacturing and enables reducing the main field sensitivity $K_{n}$ below $ 10^{-3}$ \cite{Rogacki2020}. In this way, even small deviations from multipole errors in the $10^{-4}$ range can be identified with a high resolution. We therefore focus on radial, rotating coil arrays built on a solid PCB, rather than less accurate flexible PCB technology as it was done in \cite{Arpaia2019}. In addition to transversal and torsional vibrations, an unsteady angular velocity also contributes to spurious harmonics in the acquired signals. As most systems use shafts optimized for mechanical stiffness, the largest impact is  attributed to effects in the bearings. Approaches to improve the motion of rotating coils, therefore focus on optimizing the intersection beween shaft and bearing \cite{SORTI2020164599}. Integrating the signals in time and trigger the integration periods by angular encoders yelds measurements of flux over angular position. In this way any time dependency is eliminated. The approach presented in this article is therefore based on flux increments as provided from fast digital integrators (FDIs) \cite{arpaia2006}.

\begin{figure*}[ht]
	\centering
	\includegraphics[width=0.7\linewidth]{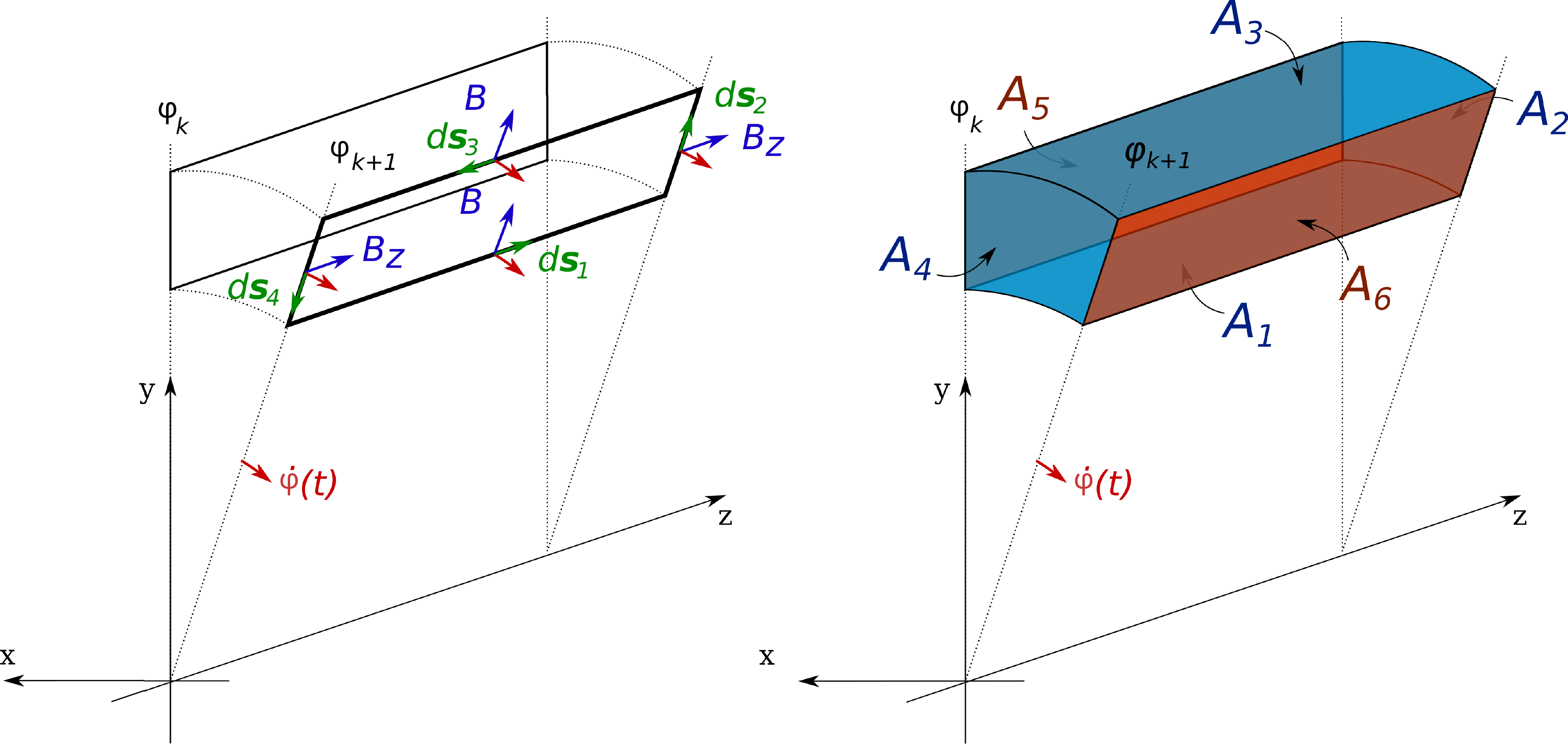}
	\caption{Reparameterisation of the integral involved in computing the flux linkage in a radial rotating coil. The physical phenomenon is based on $\bm{v}\times \bm{B}$, and thus, occurs on the blue surfaces. The integration, however, can be carried out over the red areas $A_5$ and $A_6$.}
	\label{fig:v_cross_B}
\end{figure*}

\begin{figure}
	\centering
	\includegraphics[width=0.5\linewidth]{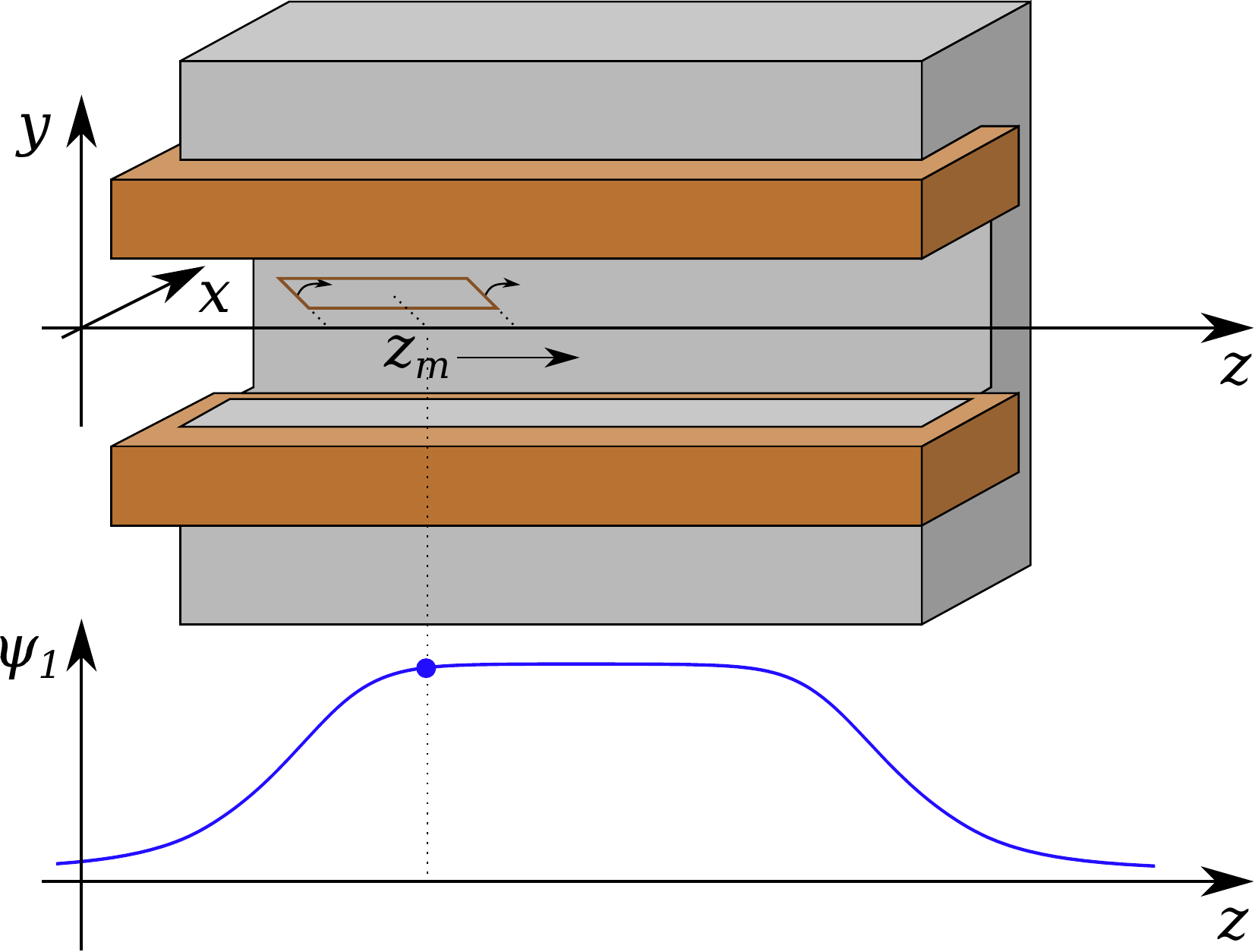}
	\caption{Measurement procedure based on longitudinally sampling the magnetic field in the air gap of the magnet with a radial rotating coil. Here $\Psi_1$ denotes the dipole component of the field.}
	\label{fig:sampling_multipoles}
\end{figure}

Moving a single wire loop surface $A$ with the velocity $\bm{v}$ in a static magnetic field $\bm{B}$, induces the voltage~\cite{Stephan2011}
\begin{align}
	U_\mathrm{ind}(t) = \int\limits_{\partial {A}(t)}\left(\bm{v}(t)\times \bm{B}\right)\cdot \mathrm{d}\bm{s},
\end{align}
by Faraday's law. Considering a radial coil as illustrated in Fig.~\ref{fig:v_cross_B}, this translates to
\begin{align}
	U_\mathrm{ind}(t) = \int\limits_{\partial {A}(t)}\left(\bm{v}(t)\times \bm{B}\right)\cdot \mathrm{d}\bm{s}\,=\,  \dot{\varphi}(t)\left(\int\limits_{S_1} r_1\bm{B}\cdot\mathrm{d}\bm{s}
	+\int\limits_{S_2} r\bm{B}\cdot\mathrm{d}\bm{s} +\int\limits_{S_3} r_3 \bm{B}\cdot\mathrm{d}\bm{s}
	+\int\limits_{S_4} r\bm{B}\cdot\mathrm{d}\bm{s}\right), 
	\label{eq:faraday_coil}\end{align}
where $S_i=S_i(t)$, $i=1,...,4$ are the boudaries of $\partial A(t)$.
The angular velocity $\dot{\varphi}(t)$ is difficult to stabilize or measure. This is why it is common practice to integrate $U_\mathrm{ind}(t)$ in time between two angular positions $\varphi_k=\varphi(t_k)$ and $\varphi_{k+1}=\varphi(t_{k+1})$. The integration period is started at $t_k$ and stopped at $t_{k+1}$ by trigger pulses generated by a rotary encoder. Integrating Eq.~\eqref{eq:faraday_coil} in time, we obtain the flux increment
\begin{align}
	\label{eq:flux}
	\delta\Phi(\varphi_{k+1},\varphi_{k},z_{\rm m})
	= \int\limits_{A_1}  \bm{B}\cdot  \mathrm{d} \bm{a} + \int\limits_{A_2}  \bm{B}\cdot  \mathrm{d} \bm{a}
	+\int\limits_{A_3} \bm{B}\cdot  \mathrm{d} \bm{a}
	+\int\limits_{A_4} \bm{B}\cdot  \mathrm{d} \bm{a}
	=-\int\limits_{A_6} \bm{B}\cdot  \mathrm{d} \bm{a}
	-\int\limits_{A_5} \bm{B}\cdot  \mathrm{d} \bm{a},
\end{align} 
where we have applied Gauss' law to the areas $A_i=A_i(\varphi_{k+1},\varphi_{k},z_{\rm m})$, $i=1,...,6$ illustrated in Fig.~\ref{fig:v_cross_B} (right) and $z_{\rm m}$ denotes the position of the center of the coil (where the measurement is performed).
For a particular axial position $z_{\rm m}$, one can obtain the total flux intercepted at position $\varphi_{k+1}$ by
\begin{align}
	\Phi(\varphi_{k+1},z_{\rm m}) = \Phi(\varphi_{1},z_{\rm m})+ \sum_{j=1}^{k} \delta\Phi(\varphi_{j+1},\varphi_{j},z_{\rm m}),
	\label{eq:summation_flux}
\end{align}
where $\Phi(\varphi_{1},z_{\rm m})$ is the flux linkage at the beginning of the summation. After a full rotation of the coil, having sampled at $K$ angular positions, a discrete Fourier transformation yields an ensemble of harmonic coefficients 
\begin{align}
	\Psi_n(z_{\rm m}) = \sum_{k=1}^{K} \Phi(\varphi_k,z_{\rm m}) \exp\left(-\frac{2\pi i}{K}(k-1) n\right), \end{align}
for  $n=1,...,N,$ where usually a sufficient approximation accuracy is already obtained for $N<20$.
Contrary to classical rotating coil measurement analysis, which treats the Fourier transform of $\Phi(\varphi,z_{\rm m})$ as raw data, we are taking the flux increments directly into account. This allows for a straight-forward treatment of transversal coil offsets, by field translation (see section \ref{sec:field_rec_dPhi_k}).
Throughout this paper we aim for a determination of the three-dimensional field distribution in the air gap of an accelerator magnet. We therefore shift a short rotation coil scanner though the longitudinal extension as illustrated in Fig.~\ref{fig:sampling_multipoles}, and measure step-wise the flux linkages for different axial positions.

\section{Lobatto Spline Fourier Expansion}\label{sec:field_rec_dPhi_k}
Consider the current-free, cylindrical domain with homogeneous permeability
\begin{equation}
D=\{(x,y,z)^\top\text{ : }x^2+y^2<R^2, z_\text{min}<z<z_\text{max} \},
\label{eq:DOI}
\end{equation}
with radius $R$ and axial extent $[z_{\rm min},z_{\rm max}]$, inside the air gap of the magnet and the cylindrical coordinates $\varrho,\varphi$ and $z$. The Laplace equation
\begin{equation}
\Delta {\phi} = 0, \label{eq:laplace}
\end{equation}
holds for the magnetic scalar potential $\bm{B}= \nabla {\phi}$ in $D$.
Exploiting cylindrical coordinates we use a discretization in the forms:
\begin{equation}
{\phi}(\varrho,\varphi,z)\approx\text{Re}\left(\sum_{m,n,k} d_{m,n,k} P_m(\varrho) e^{\iu n\varphi} b_k(z)\right), \label{eq:basis_PFB}
\end{equation}
where $\{P_m(\varrho)\}_{m=1,...,N_\varrho}$ are Lobatto shape functions \cite{higher_order_fem} scaled to $[0,R_0]$, 
$\{b_k(z)\}_{k=1,...,N_z}$ is a B-spline basis \cite{dierckx1995curve} defined by a given knot vector scaled to $[z_{\text{min}},z_{\text{max}}]$ 
and $d_{m,n,k}$ are the corresponding degrees of freedom (DoF). The radial flux density component $B_\varrho =  \partial_\varrho {\phi}$ at the domain boundary $\varrho \to R$ can be approximated by a projection in the discrete space
\begin{equation}
u(\varphi,z)= B_\varrho(\varphi,z) \approx \text{Re}\left(\sum_{n,k} u_{n,k} e^{\iu n\varphi} b_k(z)\right), \label{eq:bc}
\end{equation}
and relates to the Neumann data as a boundary condition for ${\phi}$.

The discretized solution in the form \eqref{eq:basis_PFB} for problem \eqref{eq:laplace} with BC \eqref{eq:bc} is obtained by a Galerkin approximation, leading to \cite{higher_order_fem,fem_monk}:
\begin{equation}
\bm{L}\bm{d}=\bm{P}\bm{u}, \label{eq:system}
\end{equation}
where $\bm{L}\in\mathbb{R}^{N_\varrho N_\varphi N_z\times N_\varrho N_\varphi N_z}$ is the discrete Laplace operator, $\bm{d}\in\mathbb{R}^{N_\varrho N_\varphi N_z}$ is the vector corresponding to the DoF of the scalar potential approximation, $\bm{P}\in\mathbb{R}^{N_\varrho N_\varphi N_z\times  N_\varphi N_z}$ is the extension operator and $\bm{u} \in \mathbb{R}^{N_\varphi N_z}$ are the DoF of the Neumann data. Due to the tensor-product format of the basis, the operator $\bm{L}$ can be reshaped into a block diagonal structure, making its inversion computationally affordable. Given the discrete representation of the Neumann data $\bm{u}$, we can get the field inside the domain by solving \eqref{eq:system}.

\begin{figure}
	\centering
	%
	%
	%
	%
	%
	%
	%
	%
	\includegraphics{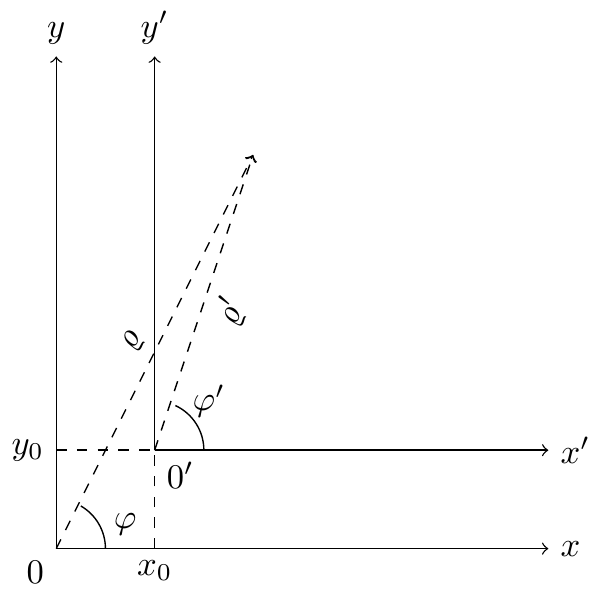}
	\caption{Coordinate transformation.}
	\label{fig:translation}
\end{figure}

In our case, however, $\bm{u}$ is an unknown quantity, and needs to be determined from flux measurements in $D$. The flux measurement process presented in Sec. \ref{sec:coil_meas} can also be constructed from Eq.~\eqref{eq:basis_PFB} by integration over the coil surface. For a single-wired coil of length $l_{\rm c}$, width $w$ and center at $(\varrho_{\rm c},\varphi(t),z_{\rm m})$, this means:
\begin{equation} \begin{split}
\Phi(\varphi,z_{\rm m})=&\int\limits_{z_{\rm m}-l_{\rm c}/2}^{z_{\rm m}+l_{\rm c}/2}\int\limits_{\varrho_{\rm c}-w/2}^{\varrho_{\rm c}+w/2} B_\varphi(\varrho,\varphi,z)\,\mathrm{d}\varrho\mathrm{d}z = \text{Re}\left(\sum\limits_{m,n,k} d_{m,n,k} \int\limits_{z_{\rm m}-l_{\rm c}/2}^{z_{\rm m}+l_{\rm c}/2}\int\limits_{\varrho_{\rm c}-w/2}^{\varrho_{\rm c}+w/2} \frac{\iu n}{\varrho} P_m(\varrho) e^{\iu n\varphi} b_k(z)\mathrm{d}\varrho\mathrm{d}z\right). 
\end{split}   \label{eq:flux1}
\end{equation}
In \eqref{eq:flux1}, the offset of the rotation axis can also be taken into account by translating the coordinate system (see Fig. \ref{fig:translation}). The origin of the reference coordinate system is denoted as $0$ and the offset of the rotation axis is denoted as $0'$ and has the coordinate $(x_0,y_0)$. The transformed coordinates are
\begin{subequations}
	\begin{align}
		\varrho(\varrho',\varphi') = & \sqrt{(\varrho'\sin\varphi'+y_0)^2+(\varrho'\cos\varphi'+x_0)^2},\label{eq:translation}\\ 
		\varphi(\varrho',\varphi') = & \text{sgn}(\varrho'\sin\varphi'+y_0)\text{acos}\left(\frac{\varrho'\cos\varphi'+x_0)}{\varrho(\varrho',\varphi')}\right). \label{eq:translation2}
	\end{align}
\end{subequations}

Equations \eqref{eq:translation} and \eqref{eq:translation2} can be used as a parametrization of the coil and Eq.~\eqref{eq:flux1} can be adapted using the change of variables in order to get the flux measured around a point $\bm{r}^0 = (x_0,y_0,z_0)^\top$. The resulting flux is linear with respect to the DoFs and can be formally represented using the measurement operator
\begin{equation}
\Phi(\varphi_l,\bm{r}^0) =  \text{Re}\left(\sum\limits_{m,n,k} d_{m,n,k} M_{l,m,n,k}(\bm{r}^0)\right), 
\label{eq:flux2}
\end{equation}
where $l = 1,...,M$. 

Note that the rotating coil is made by multiple windings, which are not placed at the same position. The real layout can thus be much more complex (see Fig.~\ref{fig:PCB_layout}) than a single-turn induction coil, for which Eq.~\eqref{eq:flux1} and Eq.~\eqref{eq:flux2} were derived. However, the real coil layout can be modelled by superimposing Eq.~\eqref{eq:flux2} for multiple windings with varying geometrical properties. The overall measurement operator can then be reshaped into an $M\times (N_\varrho N_\varphi N_z)$ matrix $\bm{M}(\bm{r}^0)$ which relates the DoF vector $\bm{d}$ to the flux measurements performed for $M$ angles and concatenated into a vector $\bm{y}\in\mathbb{R}^{M}$. Combining the operators, we get the discrete observation operator: 
\begin{equation}
\bm{S}_{\bm{r}^0}=\bm{M}(\bm{r}^0)\bm{L}^{-1}\bm{P},\end{equation}
that gives the predicted measurement for a given boundary data from the discrete space:
\begin{equation*}
	\bm{u} \underset{\bm{L}^{-1}\bm{P}}{\longrightarrow} \bm{d} \underset{\bm{M}(z,x_0,y_0)}{\longrightarrow} \bm{y}.
\end{equation*}

\section{Bayesian Inversion}\label{sec:bayesian_inversion}

In this section, we focus on the identification of the DoF vector of the field representation using the measured flux increments. In a more general setting, we consider a system described by a state which can be observed indirectly by measurements. Using the governing mathematical model, one can predict the observation for a given state. This procedure is called a forward problem \cite{bayes_inv}. Contrarily, an inverse problem refers to the estimation of a state given an observation, effectively inverting the model that maps states to observations \cite{bayes_inv}. Inverting this map to get the causing state of a typically noisy observation is in most cases an ill-posed problem.



\begin{figure}[t]
	\centering
	\includegraphics[width=0.95\textwidth]{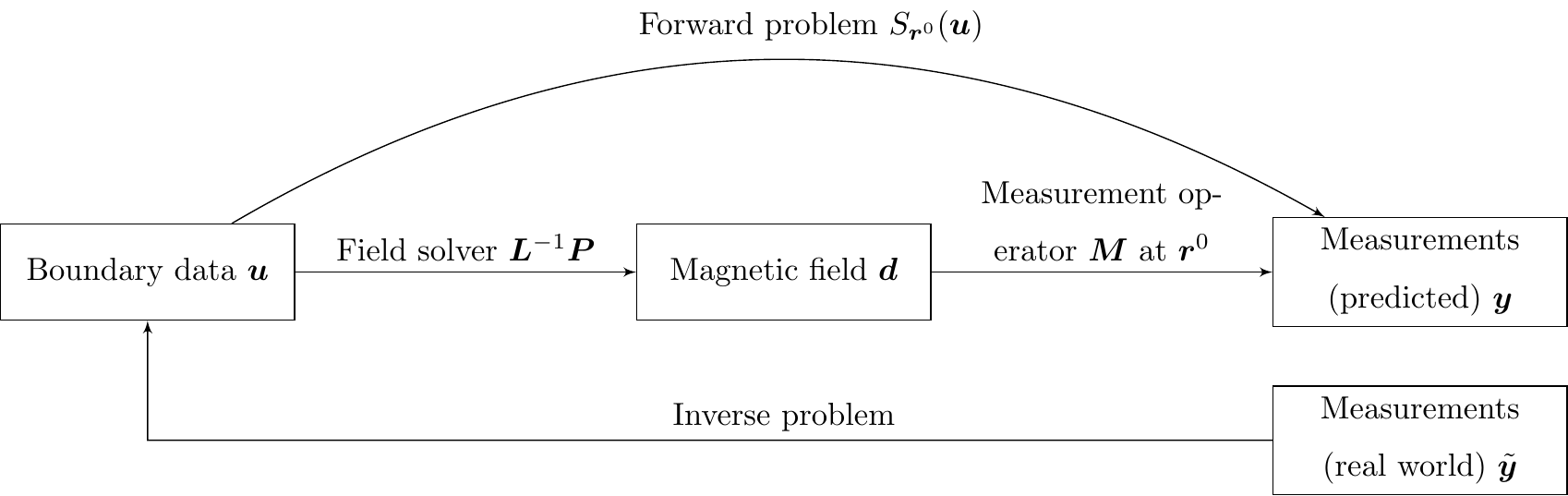}
	\caption{Forward and inverse problem.}
	\label{fig:fwd_inv}
\end{figure}

For the magnetic field reconstruction, the state of the system is the radial component of the magnetic field on the boundary, while the observations are the measured flux increments over the radial coil at different positions. The field has to fulfil Maxwell's equations inside the domain of interest, i.e. in the cylinder inside the aperture. Since the field inside the domain is uniquely described by the radial component ${B}_\varrho$ for $\varrho=R$ together with the 0 Dirichlet condition at the ends of the cylinder, the problem is well-posed if we take into account the zero field conditions at the ends of the cylinder. In this case, the forward problem consists of computing the flux increments from the boundary data by first computing the field solution (see Fig. \ref{fig:fwd_inv}) and then applying the measurement operator, denoted with $S_{\bm{r}^{0}}(\bm{u}) = \bm{S} (\bm{r}^{0}) \bm{u}$.

Considering the inverse problem, an additive Gaussian noise model is assumed for the observations. Given the forward operator $S_{\bm{r}^{0}}$ corresponding to a rotation center $\bm{r}^{0}$ and the state (boundary data) $\bm{u}$, the measurement model has the following form:
\begin{equation}
\bm{Y} = S_{\bm{r}^{0}}(\bm{u})+\bm{\epsilon}, \label{eq:obs_rv}
\end{equation}
where $\bm{\epsilon}$ is a normal distributed random variable with zero mean and covariance matrix $\bm{\Lambda}_{\rm N}$ (we write $\bm{\epsilon}\sim\mathcal{N}(\bm{0},\bm{\Lambda}_{\rm N})$). The conditional probability density function (PDF) of $\bm{Y}$ given a state is called the \textit{likelihood function} \cite[Chapter~5]{uq_inv} and can be expressed as:
\begin{equation*}
	p_{\bm{Y}|\bm{U}}(\bm{y}|\bm{u},\bm{r}^{0}) \propto \exp{\left( -\frac{1}{2}\left(S_{\bm{r}^{0}}(\bm{u})-\bm{y}\right)^\top\bm{\Lambda}_{\rm N}^{-1}\left(S_{\bm{r}^{0}}(\bm{u})-\bm{y}\right)\right)}.
\end{equation*}
In the Bayesian setup, the unknown states are also considered to be random variables, denoted in the following with $\bm{U}$. The model in \eqref{eq:obs_rv} becomes
\begin{equation}
\bm{Y} = S_{\bm{r}^{0}}(\bm{U})+\bm{\epsilon}
\end{equation}
and the goal is to update the probability density function of $\bm{U}$, by incorporating the information obtained from a new measurement $\tilde{\bm{y}}\in\mathbb{R}^M$. The distribution of the state $\bm{U}$ given a measurement is called \textit{posterior} and its probability density function $p_{\bm{U}|\bm{Y}}$ can be related to the likelihood function through the Bayes rule \cite{bayes_inv,probability_theory}
\begin{equation}
p_{\bm{U}|\bm{Y}}(\bm{u}|\bm{y},\bm{r}^{0})\propto p_{\bm{Y}|\bm{U}}(\bm{y}|\bm{u},\bm{r}^{0})p_{\bm{U}}(\bm{u}), \label{eq:bayes2}
\end{equation}
where $p_{\bm{U}}(\bm{u})$ is called the \textit{prior} and represents the PDF of the state before the observation was performed. The prior can be provided from an ideal model of the state or estimated from previous measurements. If the operator $S_{\bm{r}^0}$ is linear with respect to $\bm{u}$, $S_{\bm{r}^0}(\bm{u}) = \bm{S}({\bm{r}^0})\bm{u}$ and the prior is normally distributed with mean $\bm{u}^0$ and covariance matrix $\bm{\Sigma}$, the posterior is again Gaussian \cite{uq_inv}:
\begin{equation}
p_{\bm{U}|\bm{Y}}(\bm{u}|\bm{y},\bm{r}^{0})\propto\exp\left(-\frac{1}{2}\left(\bm{u}-\bm{\mu}'\right)^\top\bm{\Sigma'}^{-1}\left(\bm{u}-\bm{\mu}'\right)\right), \label{eq:posterior_gauss}
\end{equation}
where 
\begin{align}
	\bm{\mu}'=\bm{L}(\bm{S}^\top\bm{\Lambda}_{\rm N}^{-1} \bm{y} +\bm{\Sigma}^{-1} \bm{u}^0),\\
	\bm{\Sigma}'=\left(\bm{S}^\top\bm{\Lambda}_{\rm N}^{-1}\bm{S}+\bm{\Sigma}^{-1}\right)^{-1}.
\end{align}
\subsection{Kalman filter}
The mean and the covariance of the Gaussian posterior \eqref{eq:posterior_gauss} can be reformulated to obtain the equations of the Kalman filter \cite{bayes_inv,kalman}. Using the matrix inversion rule \cite{hager1989updating}, the posterior covariance matrix is
\begin{equation}
\bm{\Sigma'} = \left(\bm{S}^\top\bm{\Lambda}_{\rm N}^{-1}\bm{S}+\bm{\Sigma}^{-1}\right)^{-1} = \bm{\Sigma} - \bm{\Sigma}\bm{S}^\top (\bm{\Lambda}_{\rm N} + \bm{S} \bm{\Sigma} \bm{S}^\top)^{-1} \bm{S} \bm{\Sigma} = \bm{\Sigma} - \bm{K} \bm{S} \bm{\Sigma} = (\bm{I}-\bm{K} \bm{S})\bm{\Sigma},
\end{equation}
where $\bm{K}=\bm{\Sigma}\bm{S}^\top\left(\bm{\Lambda}_{\rm N}+\bm{S}\bm{\Sigma}\bm{S}^\top\right)^{-1}$ is called Kalman gain. The mean can be reformulated as
\begin{equation}
\bm{\mu}' = (\bm{I}-\bm{K} \bm{S})\bm{\Sigma}(\bm{S}^\top\bm{\Lambda}_{\rm N}^{-1} \bm{y} +\bm{\Sigma}^{-1} \bm{\mu}) = \bm{\mu} + \bm{K}( \bm{y}-\bm{S} \bm{\mu}).
\end{equation}
The previous two equations compute the mean $\bm{\mu}'$ and the covariance matrix $\bm{\Sigma}'$ given a measurement under the assumption that the prior is Gaussian. The resulting $\bm{\mu}'$ and $\bm{\Sigma}'$ can also be taken as priors for a new Bayesian update. For the magnet measurement, we have $N_{\rm T}$ radial coil measurements $\{\hat{\bm{y}}^{(k)}\}_{k=1,...,N_{\rm T}}$, each corresponding to rotation centers $\{\bm{r}^{(k)}\}_{k=1,...,N_{\rm T}}$.
Starting from a Gaussian prior $\mathcal{N}(\bm{\mu}^{(0)},\bm{\Sigma}^{(0)})$, the Kalman filter update scheme can be performed \cite{kalman}:
\begin{subequations}
	\begin{align}
		\bm{K}^{(k+1)} = & \bm{\Sigma}^{(k)}\bm{S}_{\bm{r}^{(k+1)}}^\top\left(\bm{\Lambda}_{\rm N}+\bm{S}_{\bm{r}^{(k+1)}}\bm{\Sigma}^{(k)}\bm{S}_{\bm{r}^{(k+1)}}^\top\right)^{-1},\\
		\bm{\mu}^{(k+1)} =  &  \bm{\mu}^{(k)}+\bm{K}^{(k+1)}\left(\hat{\bm{y}}^{(k+1)}-\bm{S}_{\bm{r}^{(k+1)}}\bm{\mu}^{(k)}\right),\\
		\bm{\Sigma}^{(k+1)} = &\left(\bm{I}-\bm{K}^{(k+1)} \bm{S}_{\bm{r}^{(k+1)}}\right)\bm{\Sigma}^{(k)}.
	\end{align}
\end{subequations}

\section{Example: Magnetic Measurements of a Corrector Dipole}\label{sec:meas_corrector_dipole}
\begin{figure}
	\centering
	\includegraphics[width=0.8\linewidth]{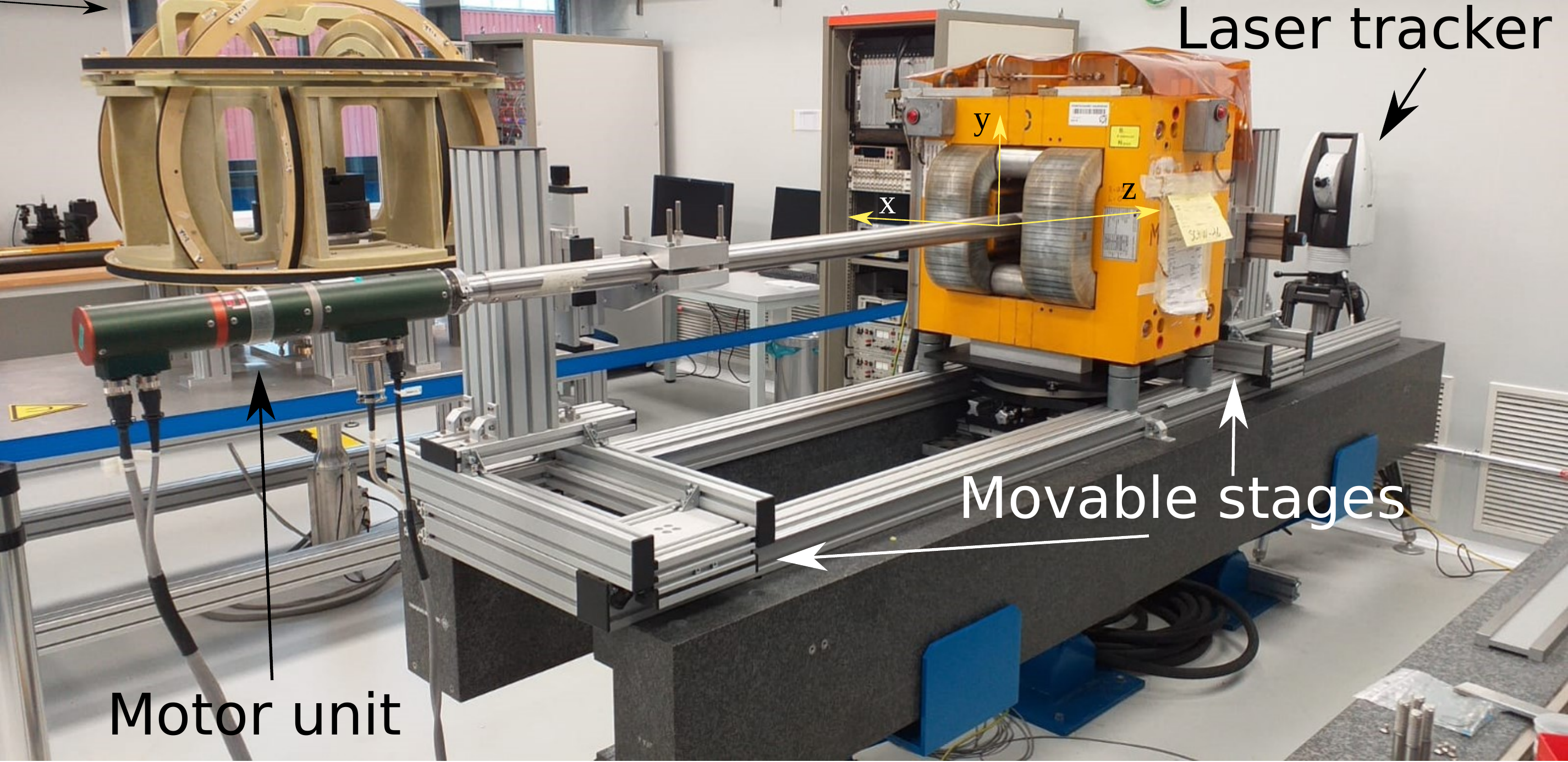}
	\caption{Corrector dipole magnet (yellow) installed on the measurement bench. The motor unit is hosting a stepper motor as well as a rotary encoder.}
	\label{fig:measurement_setup}
\end{figure}

\begin{figure}[t]
	\centering
	\begin{minipage}{0.47\textwidth}
		\centering
		\includegraphics[width=\linewidth]{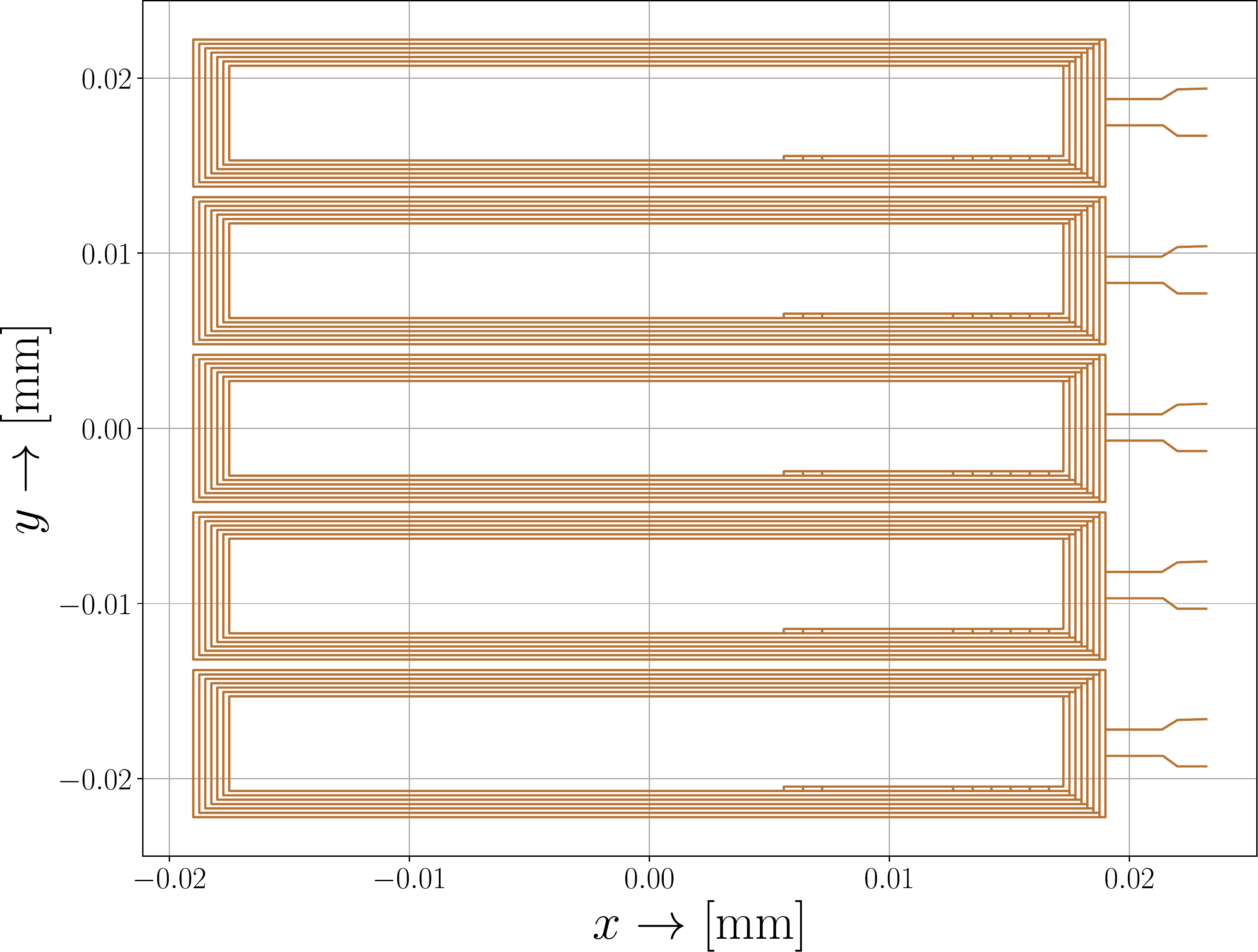}
		\caption{The PCB coil array. Only one outer coil and the central coil were used to compensate for the main dipole component.}
		\label{fig:PCB_layout}
	\end{minipage}\hfill
	\begin{minipage}{0.47\textwidth}
		\centering
		\includegraphics[width=\linewidth]{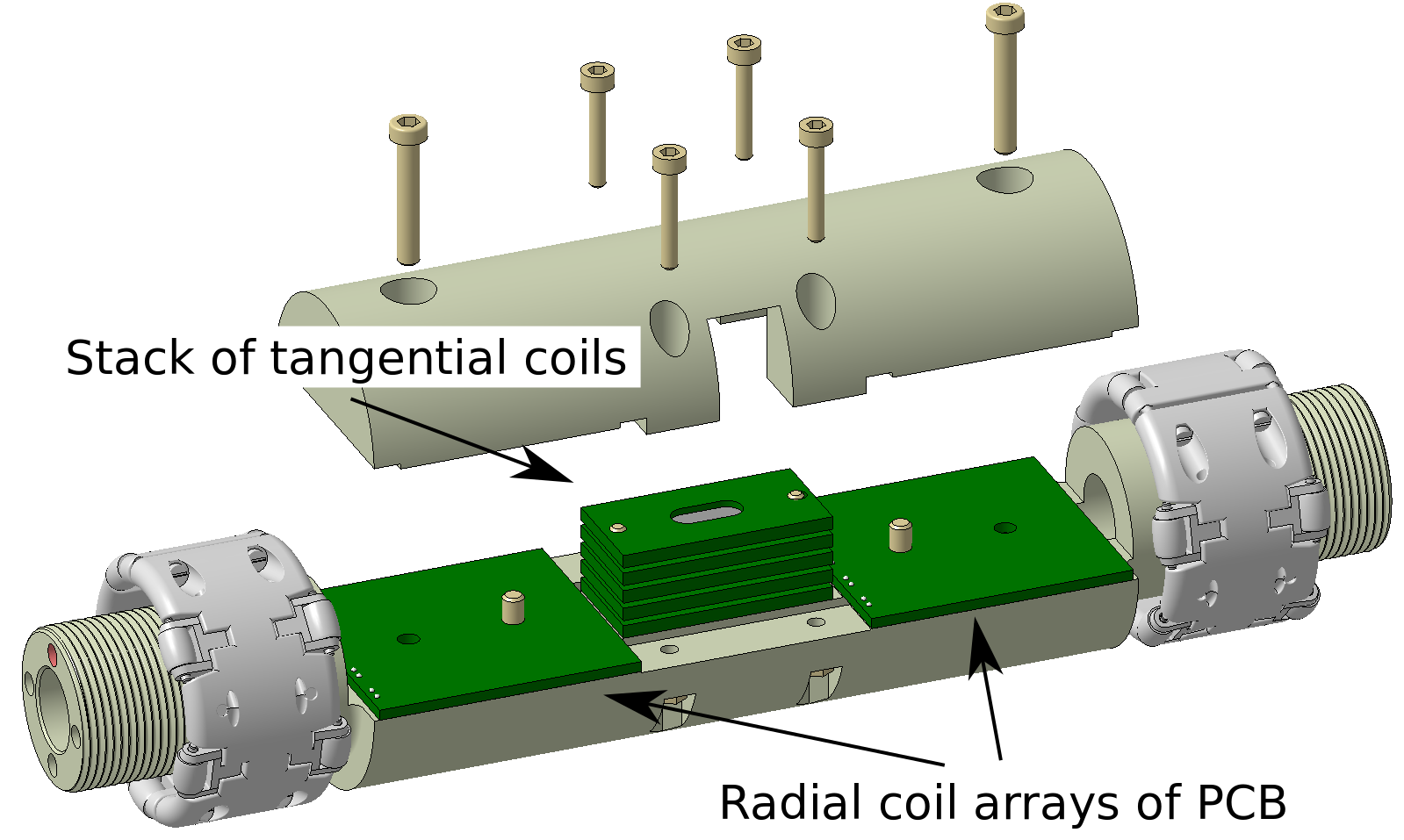}
		\caption{The shaft hosting the PCB array. There are multiple radial and tangential coils mounted on this sensor. For the tests presented in this article, only one of the radial coil arrays was used.}
		\label{fig:shaft}
	\end{minipage}
\end{figure}

Consider again the current-free, cylindrical domain from \eqref{eq:DOI} with homogeneous permeability.
Our approach was used to measure the longitudinal field profile in a corrector dipole magnet. The measurement setup is shown in Fig.~\ref{fig:measurement_setup}. 
The shaft, fixing the rotating coil array, is mounted in a tube, which is held by two alignment stages. Moving the alignment stages longitudinally, the sensor position is modified between the measurements. 
This is done by hand and requires a reasonable amount of time. In this way, 175 positions were scanned by the sensor, which took around 4 hours. The proposed strategy is designed to continue learning from more data whenever it is available. It is expected that the uncertainties in the local field reconstruction converge to a lower limit when feeding the algorithm with more data. At each position, the sensor was rotated by 20 turns and an FDI was used to acquire 512 flux increments per turn. The rotational center as well as the coil's rotational axis at each position were measured with a laser tracker targeting a retro-reflector which is mounted on the shaft's end. The coil array is built on a PCB (see Fig.~\ref{fig:PCB_layout}). It includes 5 coils made out of 7 turns on 18 layers, which are identically designed. Table~\ref{tab:geometry_sensor} gives the geometric parameters of the 5 coils.
\begin{table}
	\centering
	\begin{tabular}{ ccccc } 
		outer length & inner length & outer width & inner width & effective surface\\
		\hline
		38 mm & 35 mm & 8.4 mm & 5.4 mm & 0.0317 mm$^2$\\
		\hline
	\end{tabular}
	\caption{Designed geometric properties of the coils on the PCB. The effective surface is calculared by $\overline{w}\times \overline{l}\times\text{turns}\times\text{layers}$, whereas $\overline{w}$ and $ \overline{l}$ are averaged length and width over turns.}
	\label{tab:geometry_sensor}
\end{table}
The PCB in use is standardized measurement equipment at CERN and usually used for quadrupole measurements using a compensation scheme which employs 4 of the 5 coils. However, for this test, only one of the outer coils and the central coil are used and the dipole compensation scheme is applied, i.e. the central and outer coils are connected in counter-series to suppress the main dipole component. The shaft is illustrated in Fig.~\ref{fig:shaft}. The signal-to-noise ratio is proportional to the coil area. However, larger coils yield less localized measurement kernels. For local field measurements a trade-off between signal to noise ratio and localization has to be found. The sensor applied in this work has a length of 36 mm, yielding a high sensitivity field distributions with longitudinal frequencies below $\sim 0.027 $ [1/mm$^2$]. For steeper roll offs requiring higher frequencies, short coils or superpositions as mentioned in \cite{Arpaia2019} can be used. 

\begin{figure}
	\centering
	\includegraphics[width=0.3\linewidth]{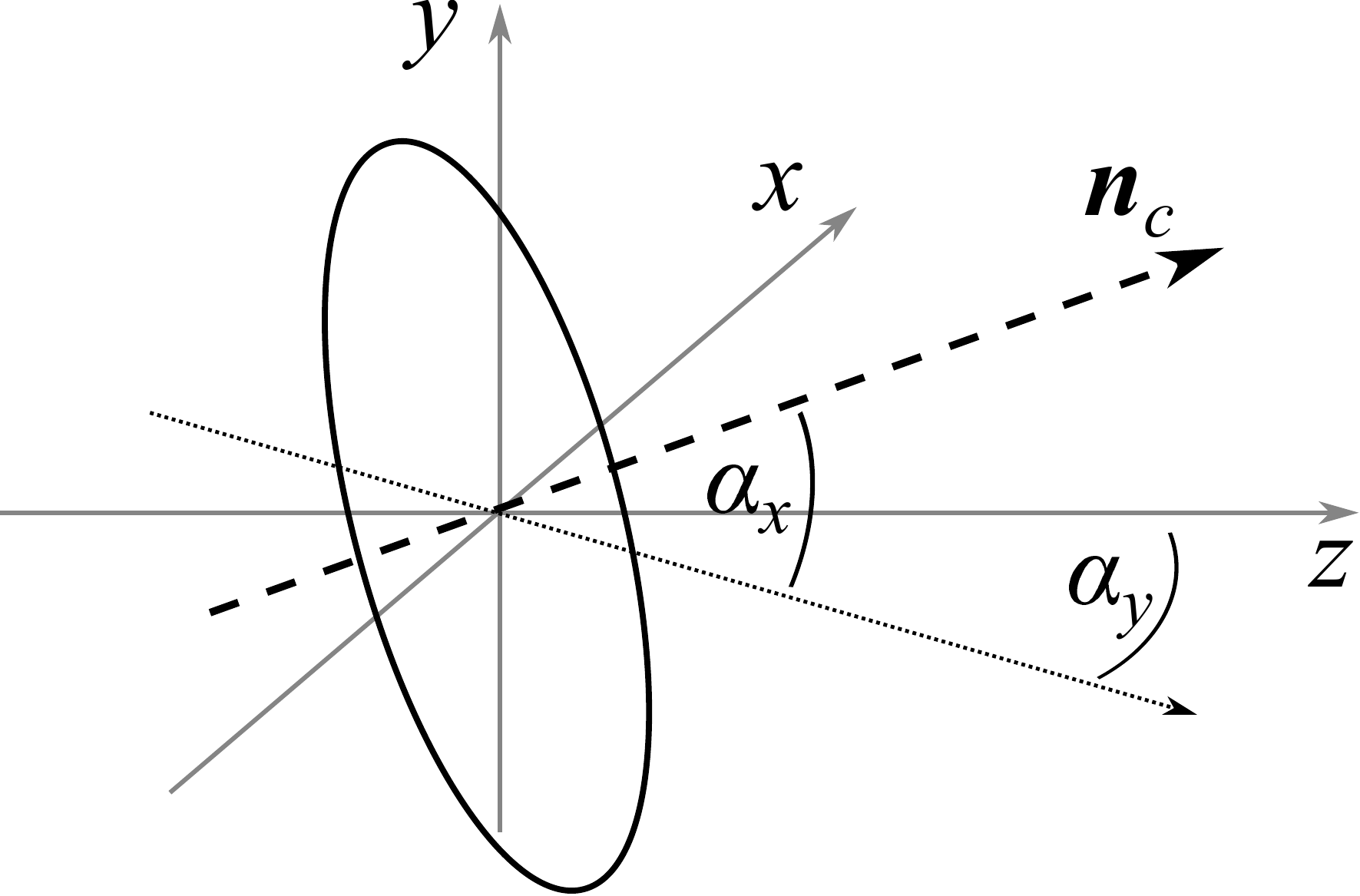}
	\caption{Definition of the angular misalignment $\alpha_x$ and $\alpha_y$.}
	\label{fig:orientation_rotation}
\end{figure}

As our field reconstruction is based on shifting a short rotating-coil sensor through the air gap, sensor alignment is critical for the quality of the field reconstruction. 
The position of a retroreflector mounted on the shaft was measured with an absolute laser tracker. Mounting the reflector off centered allows to estimate the center and the axis of rotation. The accuracy of the device is in the range of 10 um. The known coil alignment can then be accommodated in our model similar to the transversal offset. We therefore introduce the unit vector:
\begin{equation}
\bm{n}_{\rm c} = \left(\begin{array}{c}
\sin\alpha_y\cos\alpha_x \\
-\sin\alpha_x \\
\cos\alpha_y\cos\alpha_x
\end{array}\right)
\end{equation}
with $\alpha_x$ and $\alpha_y$ as illustrated in Fig.~\ref{fig:orientation_rotation}. The flux $\Phi(\varphi_l,z_m)$ is then computed for a rotation around $\bm{n}_c$.

\begin{figure}
	\centering
	\begin{subfigure}[b]{\linewidth}
		\centering
		\includegraphics{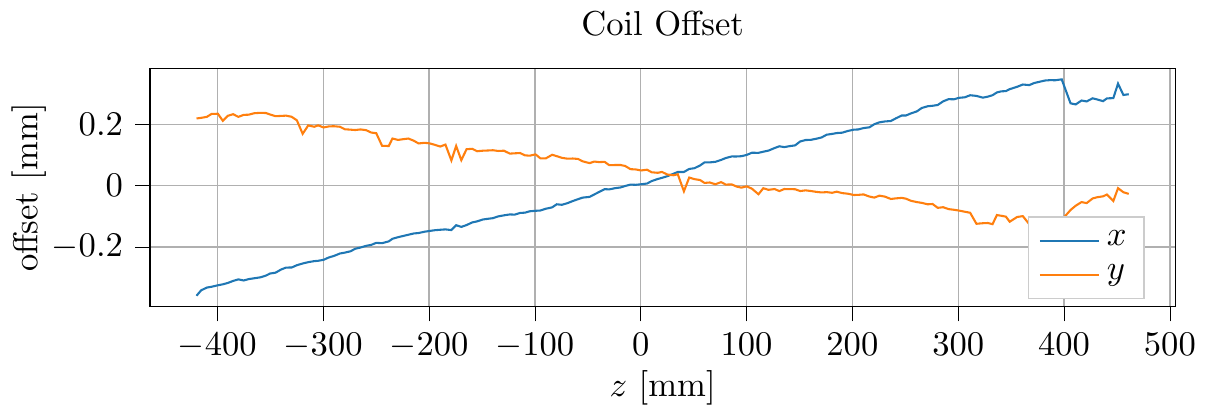}
	\end{subfigure}  
	\hfill
	\begin{subfigure}[b]{\linewidth}
		\centering
		\includegraphics{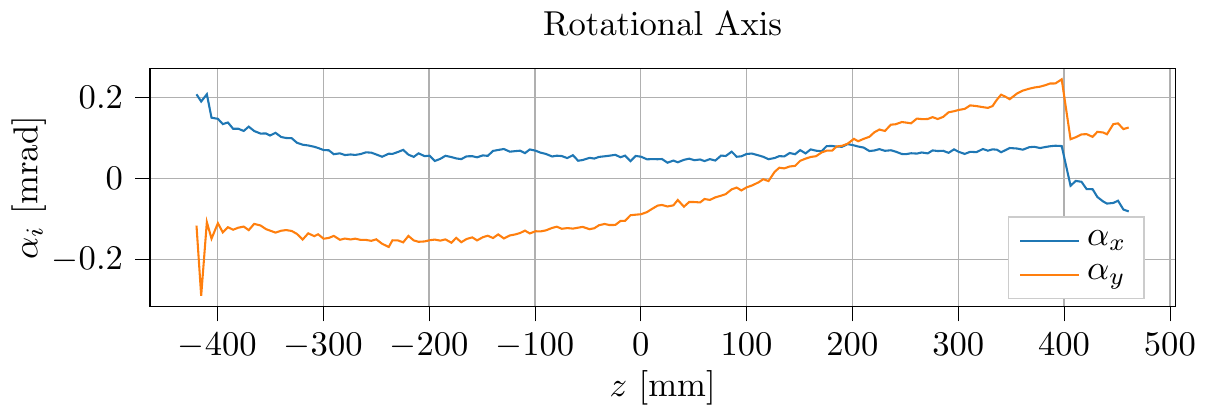}
	\end{subfigure}
	\caption{Sensor alignment over the longitudinal axis. Top: transversal offset of the rotational center. Bottom: angular misalignment of the rotational axis.}
	\label{fig:coil_orientation_and_position}
\end{figure}

As illustrated in Fig.~\ref{fig:coil_orientation_and_position}, the coil offset was kept below $0.3$ mm and the rotational axis diverged less than than 0.25 mrad over the full length of the scan. Note that the possibility for such precise alignment is not always available, especially for long magnets that cover a larger part of the tube. However, as long as it is possible to access the coil from one side by means of a laser tracker, misalignments can be measured and included in our post-processing.

\begin{figure}[t]
	\centering
	\begin{subfigure}[b]{\linewidth}
		\centering
		\includegraphics{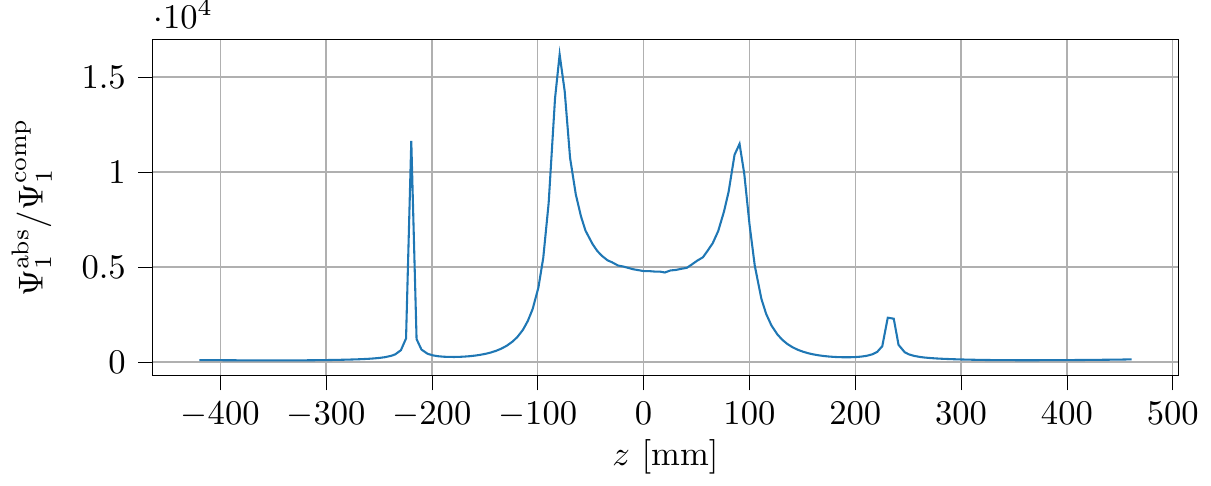}
	\end{subfigure}  
	\hfill
	\begin{subfigure}[b]{\linewidth}
		\centering
		\includegraphics{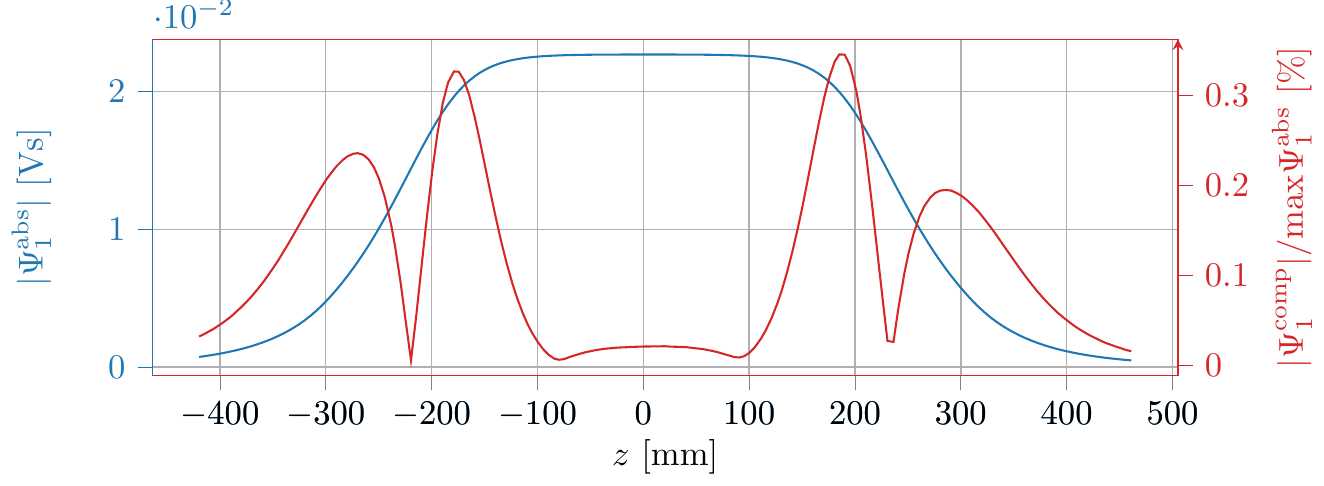}
	\end{subfigure}
	\caption{Compensation ratio over the length of the magnet. The compensation scheme is based on the scaling of $\sim\varrho^{|n|}$ over $\varrho$ and thus only applies when $\bm{B}$ is homogeneous along $z$. 
		As expected from the geometric properties of absolute and compensated coil, the compensation ratio \eqref{eq:Rcomp} of inside the magnet reaches values of $\sim 5000$.}
	\label{fig:compensation_ratio}
\end{figure}

\begin{figure}[h!]
	\begin{subfigure}[]{0.5\linewidth}
		\includegraphics{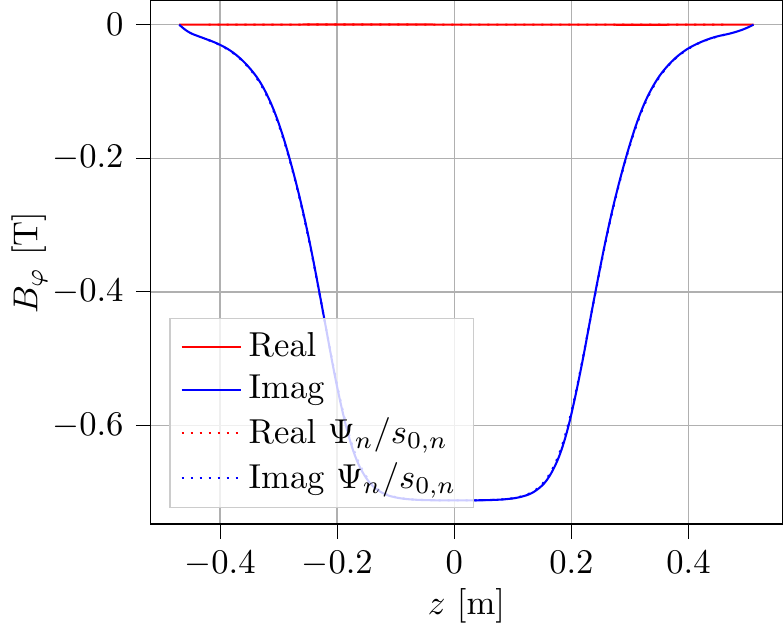}
		\caption{Dipole component}
	\end{subfigure}
	\begin{subfigure}[]{0.49\linewidth}
		\includegraphics{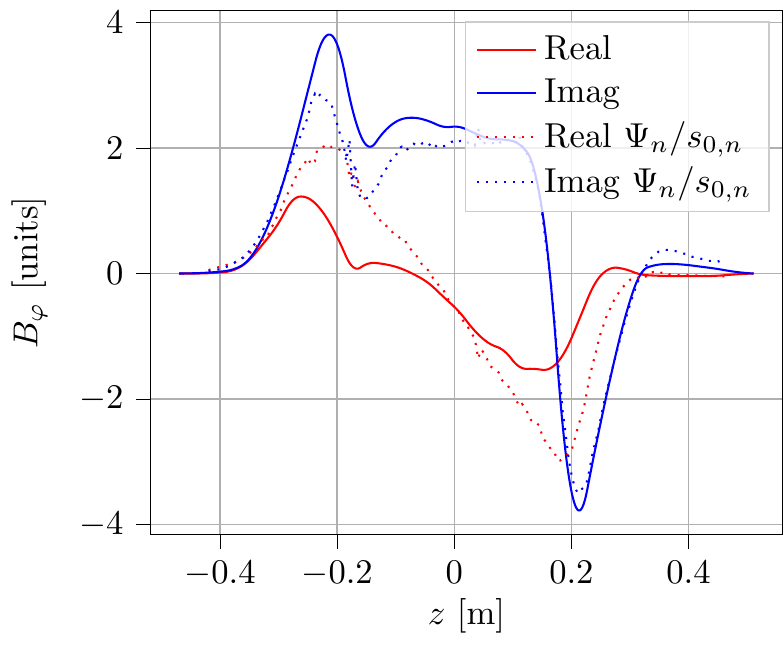}
		\caption{Quadrupole component}
	\end{subfigure} 
	\begin{subfigure}[]{0.5\linewidth}
		\includegraphics{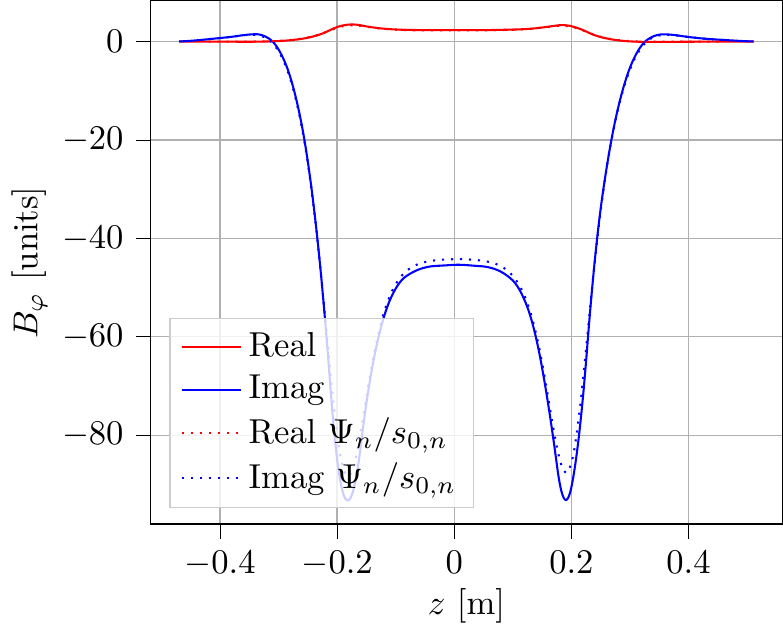} 
		\caption{Sextupole component}
	\end{subfigure}
	\begin{subfigure}[]{0.49\linewidth}
		\includegraphics{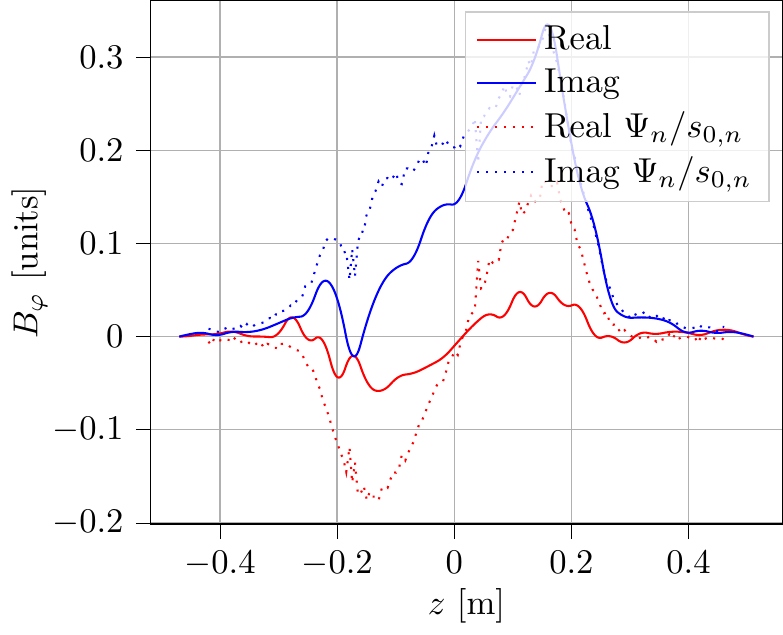}
		\caption{Octupole component}
	\end{subfigure} 
	\centering
	\begin{subfigure}[]{0.5\linewidth}
		\includegraphics{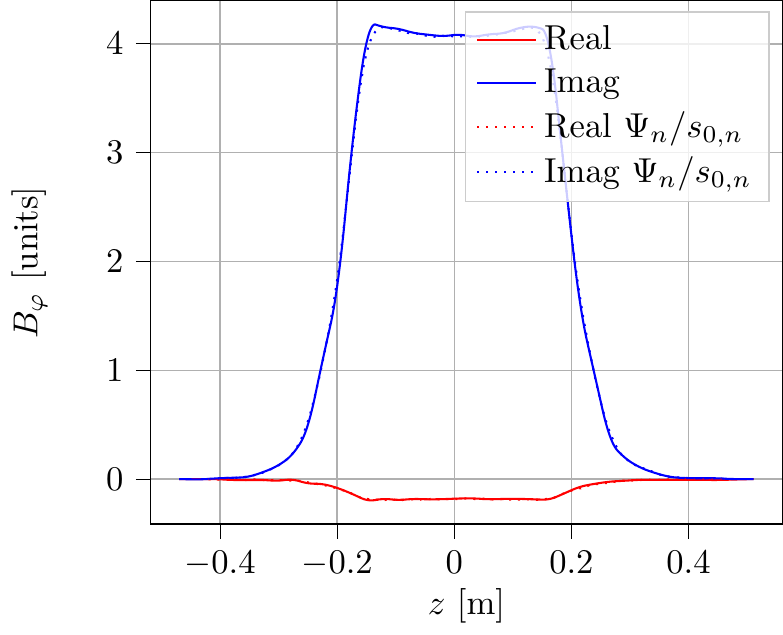}
		\caption{Decapole component}
	\end{subfigure}
	
	\caption{Multipole components of $B_\varphi$ at $R=0.018$ m. The higher order components are represented in units of 10 000 with respect to the main field component. The solid lines show the reconstruction with Bayesian inversion. The dashed lines represent the naive reconstruction.}
	\label{fig:image2}
\end{figure}

To quantify the main field suppression we look at the compensation ratio (in Fig.~\ref{fig:compensation_ratio})
\begin{equation}
R_{\rm comp}(z) = \dfrac{\Psi_{1}^{{\rm abs}}(z)}{\Psi_{1}^{{\rm comp}}(z)}, \label{eq:Rcomp}
\end{equation}
where $\Psi_{1}^{{\rm comp}}(z)$ are $\Psi_{1}^{{\rm abs}}(z)$ are the dipole components obtained with and without compensation, respectively. A large compensation ratio can be traced back to matching geometric properties of the coils involved. As expected, using two coils build on a single solid PCB, $R_{\rm comp}$ reaches values as high as $5000$ in the homogeneity region of the magnet. The compensation scheme is based on the scaling of $\sim\rho^{|n|}$ over $\rho$ and thus performs badly in the fringe field region, where $R_{\rm comp}$ drops to $250$. Interestingly, it reaches a local maximum in the fringe field where $\Psi_{1,{\rm comp}}$ evolves linearly over the coil's surface.
In Fig.~\ref{fig:image2} we illustrate the dominant field harmonics expanded in $B_\varphi$, over $z$, starting from the center. For comparison, we include the ''naive'' approach, which divides the measured flux by the integrated sensitivity factors $s_{0,n}$:
\begin{equation}
s_{0,n} = l\frac{\left(\varrho_{\rm c}+w/2\right)^{|n|}-\left(\varrho_{\rm c}-w/2\right)^{|n|}}{|n|},
\end{equation}
for a coil rotating of length $l$, width $w$ and the center at $\varrho = \varrho_\text{c}$.
All the harmonics are evaluated for a radius $\varrho = 0.018$~m. For this relatively small radius, most of the discrepancy between the naive approach and the reconstruction by field-based Bayesian inversion is attributed to the convolution of $B_\varphi$ over the area of the coil.\\

\begin{figure}[h!]
	\begin{subfigure}[]{0.5\linewidth}
		\includegraphics{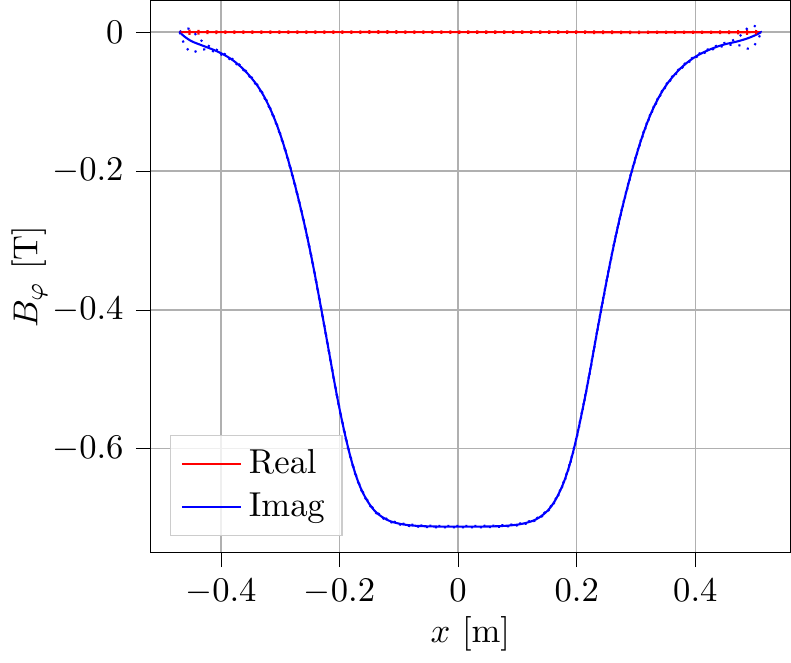}
		\caption{Dipole component}
	\end{subfigure}
	\begin{subfigure}[]{0.49\linewidth}
		\includegraphics{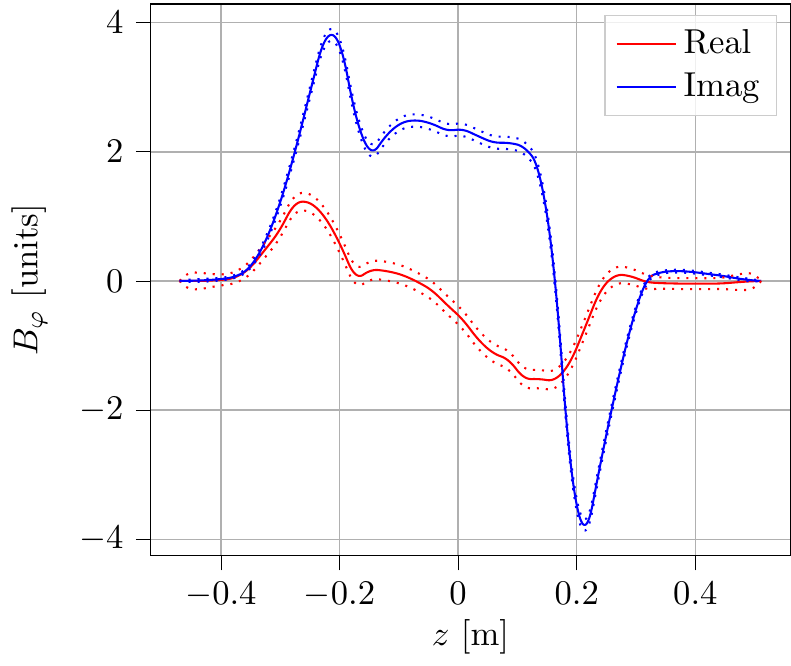}
		\caption{Quadrupole component}
	\end{subfigure} 
	\begin{subfigure}[]{0.5\linewidth}
		\includegraphics{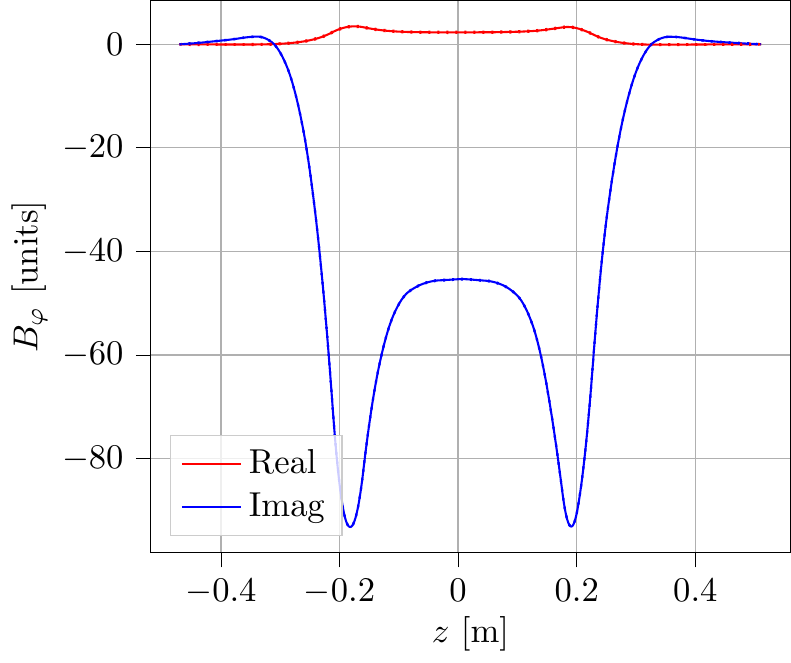}
		\caption{Sextupole component}
	\end{subfigure}
	\begin{subfigure}[]{0.49\linewidth}
		\includegraphics{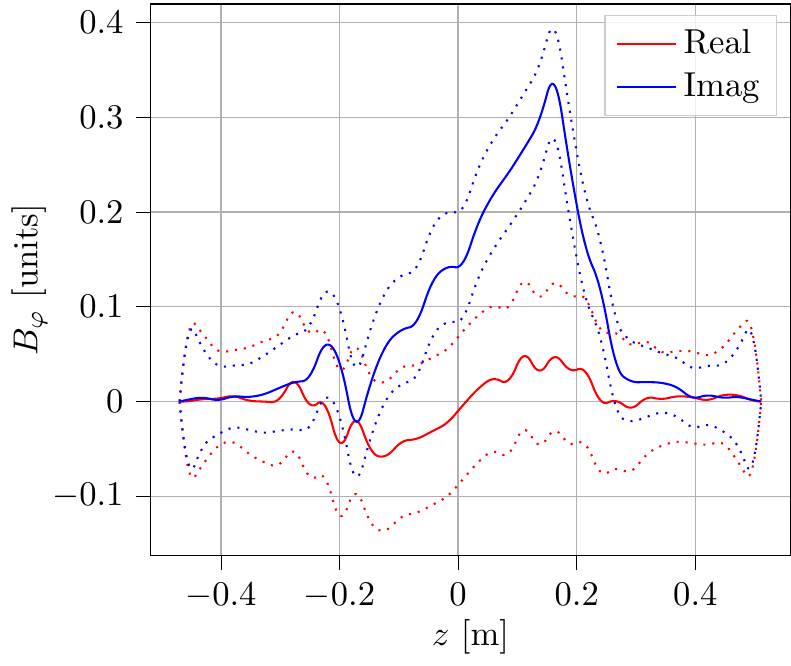}
		\caption{Octupole component}
	\end{subfigure} 
	\centering
	\begin{subfigure}[]{0.5\linewidth}
		\includegraphics{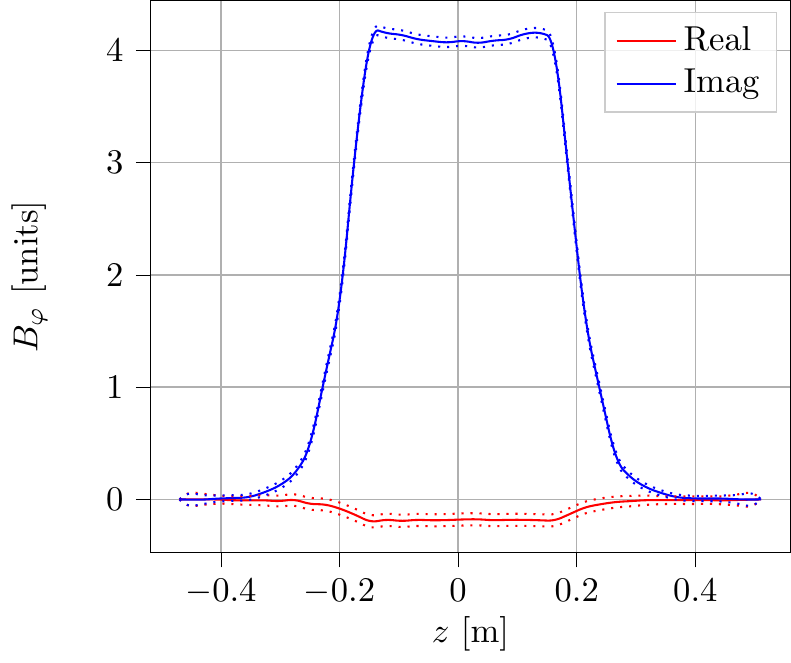}
		\caption{Decapole component}
	\end{subfigure}

	\caption{Mean $B_\varphi$ multipole components (solid lines) and $\pm2\sigma$ intervals (dashed lines).}
	\label{fig:image3}
\end{figure}

In Fig.~\ref{fig:image3} we include the $\pm 2\sigma$ credible intervals for the highest multipole coefficients. The uncertainties reach the maxima at the domain's ends. This uncertainty originates from a limited accessible $z$ interval due to the maximum displacement of the alignment stages. Future measurement campaigns will be carried out on longer test benches to circumvent this issue. The $2\sigma$ credible intervals are below 1 unit of 10000 with respect to the maximum field, for all the multipole components.\\

\begin{figure}
	\centering
	\includegraphics{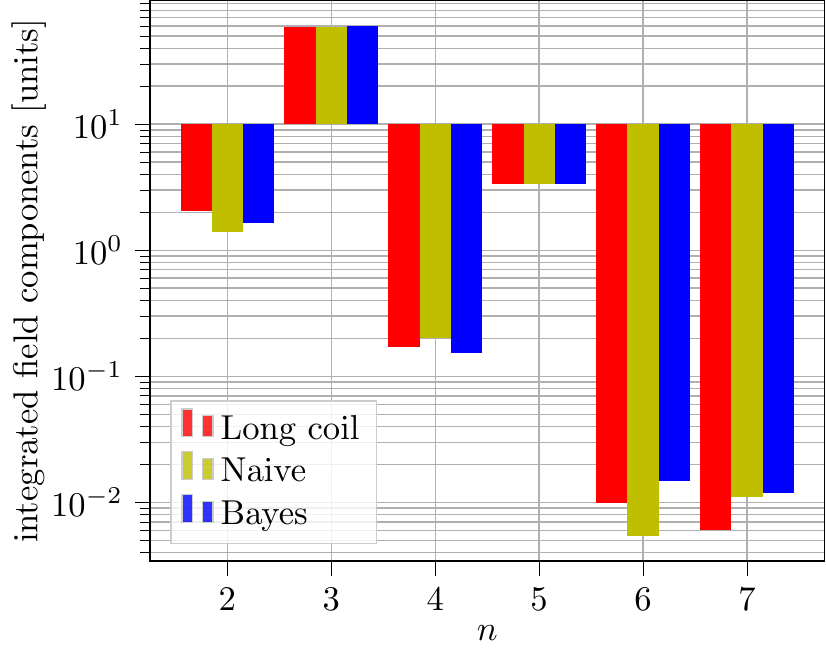}
	\caption{Comparing the field reconstruction to integrated field measurements. Here we plot the absolute of the Fourier expansion coefficients $c_n(r=18\text{mm})$ in units of 10 000 with respect to the integrated dipole field.}
	\label{fig:normal_skew_integral}
\end{figure}

The main harmonic contributions for the integral and local fields are sextupole and decapole. Quadupole and octupole component are in the small unit range and below. The differences between the two field reconstructions are visible mostly in the fringe field, where the field evaluated nonlinearly over the coil surface. Moreover, the analysis presented in this paper is capable to handle the measured sensor position and therefore correctly includes the effect of sensor misalignment. Classical feed-down corrections fail to correct misalignments in the fringe fields. This effect is clearly visible in the difference with respect to the naive approach for the quadrupole component.
Local field distributions can be validated by integrated field measurements (Fig.~\ref{fig:normal_skew_integral}). In this case, a single stretched wire measurement was used to benchmark the analysis. This comparision is shown in Figure 13. Sextupole and decapole component are match within subunit level. The differences in even ordered harmonics are slightly higher. These multipoles are highly sensitive to  alignment errors between the systems. 
\clearpage

\section{Conclusion}

Conventional approaches to measure the magnetic flux density distribution in particle accelerator magnets are based on fitting the results of rotating coil measurements to general field solutions of the Laplace equation in cylindrical coordinates. Difficulties arise due to the sensor's blind eye towards field components with steep variations along $z$, as well as the failure of classical sensor offset corrections in the fringe field region. 
For this reason we presented an approach that uses a special B-spline basis for the longitudinal variation of the transverse multipoles in order to maintain the measurement's locality. The fullfillments of Maxwell's equations is achieved via a standard Galerkin approach on the proposed discretization. Recovering the DoF based on the measurement data yields an inverse problem which is solved by Bayesian inversion. This not only enables us to infer measurement data from arbitrary measurement sources, it also paves the way for the quantification of uncertainties in the field reconstruction from noisy measurement data. 
As a particularity of Bayesian inversion, the DoF of the boundary data are represented as random variables and the goal is to characterize their joint distribution given the measured data set.
If the forward problem is linear and the noise distribution as well as the prior are assumed to be Gaussian, the Kalman filter equations can be derived. This gives the mean and the covariance of the DoF.
One significant advantage of the proposed method is that it also provides confidence information for the coefficients of the boundary data basis representation, which is incorporated into the covariance matrix of the posterior.
One further advantage is the fact that any linear measurement operator can be used. In the case of nonlinear operators, the Kalman filter is no longer applicable, however, several alternatives are presented in the literature \cite{uq_inv,RTO,cde_article}.
As a comparison, the method is validated against the classical naive reconstruction method.
The approach competes to local field measurement provided by 3D Hall probes. In view of the autors, the proposed method using a rotating coil measurement system is superior, due to the high maintainance needed to provide Hall probe measurements within the same accuracy. A detailed comparison between the approaches is currently in preparation.

\section*{Acknowledgment}
The authors would like to thank Prof. Dr.-Ing. Ulrich R\"omer from the Technische Universit\"at Braunschweig for fruitful discussions on inverse problems and uncertainty quantification.

The work of Ion Gabriel Ion and Melvin Liebsch is supported by the Graduate School CE within the Centre for
Computational Engineering at Technische Universit\"at Darmstadt.
The work of D. Loukrezis is supported by the German Ministry for Education and Research (BMBF) via the research contract 05K19RDB.

\bibliography{main_template_one_file}

\begin{thebibliography}{10}
\expandafter\ifx\csname url\endcsname\relax
  \def\url#1{\texttt{#1}}\fi
\expandafter\ifx\csname urlprefix\endcsname\relax\def\urlprefix{URL }\fi
\expandafter\ifx\csname href\endcsname\relax
  \def\href#1#2{#2} \def\path#1{#1}\fi

\bibitem{Kazantseva_2019}
E.~Kazantseva, H.~Weick, M.~Berz, K.~Makino, O.~Boine-Frankenheim,
  \href{http://dx.doi.org/10.1016/j.nima.2019.04.086}{Accurate taylor transfer
  maps for large aperture iron dominated magnets used in charged particle
  separators and spectrometers}, Nuclear Instruments and Methods in Physics
  Research Section A: Accelerators, Spectrometers, Detectors and Associated
  Equipment 935 (2019) 56–64.
\newblock \href {https://doi.org/10.1016/j.nima.2019.04.086}
  {\path{doi:10.1016/j.nima.2019.04.086}}.
\newline\urlprefix\url{http://dx.doi.org/10.1016/j.nima.2019.04.086}

\bibitem{PUGNAT2020164350}
T.~Pugnat, B.~Dalena, A.~Simona, L.~Bonaventura,
  \href{http://www.sciencedirect.com/science/article/pii/S0168900220307476}{Computation
  of beam based quantities with 3d final focus quadrupoles field in circular
  hadronic accelerators}, Nuclear Instruments and Methods in Physics Research
  Section A: Accelerators, Spectrometers, Detectors and Associated Equipment
  978 (2020) 164350.
\newblock \href {https://doi.org/https://doi.org/10.1016/j.nima.2020.164350}
  {\path{doi:https://doi.org/10.1016/j.nima.2020.164350}}.
\newline\urlprefix\url{http://www.sciencedirect.com/science/article/pii/S0168900220307476}

\bibitem{SIMONA201933}
A.~Simona, L.~Bonaventura, T.~Pugnat, B.~Dalena,
  \href{http://www.sciencedirect.com/science/article/pii/S0010465519300359}{High
  order time integrators for the simulation of charged particle motion in
  magnetic quadrupoles}, Computer Physics Communications 239 (2019) 33 -- 52.
\newblock \href {https://doi.org/https://doi.org/10.1016/j.cpc.2019.01.018}
  {\path{doi:https://doi.org/10.1016/j.cpc.2019.01.018}}.
\newline\urlprefix\url{http://www.sciencedirect.com/science/article/pii/S0010465519300359}

\bibitem{VENTURINI1999387}
M.~Venturini, A.~J. Dragt,
  \href{http://www.sciencedirect.com/science/article/pii/S0168900298015186}{Accurate
  computation of transfer maps from magnetic field data}, Nuclear Instruments
  and Methods in Physics Research Section A: Accelerators, Spectrometers,
  Detectors and Associated Equipment 427~(1) (1999) 387 -- 392.
\newblock \href {https://doi.org/https://doi.org/10.1016/S0168-9002(98)01518-6}
  {\path{doi:https://doi.org/10.1016/S0168-9002(98)01518-6}}.
\newline\urlprefix\url{http://www.sciencedirect.com/science/article/pii/S0168900298015186}

\bibitem{simona2020}
A.~Simona, \href{http://tuprints.ulb.tu-darmstadt.de/11687/}{Numerical methods
  for the simulation of particle motion in electromagnetic fields}, Ph.D.
  thesis, Politecnico di Milano, Technische Universit{\"a}t Darmstadt, Milano,
  Darmstadt (March 2020).
\newline\urlprefix\url{http://tuprints.ulb.tu-darmstadt.de/11687/}

\bibitem{bayes_inv}
H.~G. {Matthies}, E.~{Zander}, B.~V. {Rosi{\'c}}, A.~{Litvinenko}, O.~{Pajonk},
  {Inverse Problems in a Bayesian Setting}, arXiv e-prints (2015)
  arXiv:1511.00524\href {http://arxiv.org/abs/1511.00524}
  {\path{arXiv:1511.00524}}.

\bibitem{Jain:1246517}
A.~K. Jain, \href{https://cds.cern.ch/record/1246517}{{Harmonic coils}} (1998).
\newblock \href {https://doi.org/10.5170/CERN-1998-005.175}
  {\path{doi:10.5170/CERN-1998-005.175}}.
\newline\urlprefix\url{https://cds.cern.ch/record/1246517}

\bibitem{SORTI2020164599}
S.~Sorti, C.~Petrone, S.~Russenschuck, F.~Braghin,
  \href{https://www.sciencedirect.com/science/article/pii/S0168900220309967}{A
  magneto-mechanical model for rotating-coil magnetometers}, Nuclear
  Instruments and Methods in Physics Research Section A: Accelerators,
  Spectrometers, Detectors and Associated Equipment 984 (2020) 164599.
\newblock \href {https://doi.org/https://doi.org/10.1016/j.nima.2020.164599}
  {\path{doi:https://doi.org/10.1016/j.nima.2020.164599}}.
\newline\urlprefix\url{https://www.sciencedirect.com/science/article/pii/S0168900220309967}

\bibitem{Rogacki2020}
P.~Rogacki, L.~Fiscarelli, S.~Russenschuck, K.~Hameyer,
  \href{https://jsss.copernicus.org/articles/9/99/2020/}{Development of a
  rotating-coil scanner for superconducting accelerator magnets}, Journal of
  Sensors and Sensor Systems 9~(1) (2020) 99--107.
\newblock \href {https://doi.org/10.5194/jsss-9-99-2020}
  {\path{doi:10.5194/jsss-9-99-2020}}.
\newline\urlprefix\url{https://jsss.copernicus.org/articles/9/99/2020/}

\bibitem{Arpaia2019}
P.~Arpaia, G.~Caiafa, S.~Russenschuck,
  \href{https://doi.org/10.1038/s41598-018-37371-3}{A rotating-coil
  magnetometer for scanning transversal field harmonics in accelerator
  magnets}, Scientific Reports 9~(1) (Feb. 2019).
\newblock \href {https://doi.org/10.1038/s41598-018-37371-3}
  {\path{doi:10.1038/s41598-018-37371-3}}.
\newline\urlprefix\url{https://doi.org/10.1038/s41598-018-37371-3}

\bibitem{arpaia2006}
P.~Arpaia, L.~Bottura, P.~Cimmino, D.~Giloteaux, A.~Masi, J.~García~Pérez,
  G.~Spiezia, L.~Walckiers, A fast digital integrator for magnetic field
  measurements at cern, 2006, pp. 67 -- 71.
\newblock \href {https://doi.org/10.1109/IMTC.2006.328175}
  {\path{doi:10.1109/IMTC.2006.328175}}.

\bibitem{Stephan2011}
S.~Russenschuck,
  \href{https://onlinelibrary.wiley.com/doi/abs/10.1002/9783527635467.ch4}{Maxwell's
  Equations and Boundary Value Problems in Magnetostatics}, John Wiley \& Sons,
  Ltd, 2011, Ch.~4, pp. 137--185.
\newblock \href
  {http://arxiv.org/abs/https://onlinelibrary.wiley.com/doi/pdf/10.1002/9783527635467.ch4}
  {\path{arXiv:https://onlinelibrary.wiley.com/doi/pdf/10.1002/9783527635467.ch4}},
  \href {https://doi.org/10.1002/9783527635467.ch4}
  {\path{doi:10.1002/9783527635467.ch4}}.
\newline\urlprefix\url{https://onlinelibrary.wiley.com/doi/abs/10.1002/9783527635467.ch4}

\bibitem{higher_order_fem}
P.~Solin, K.~Segeth, I.~Dolezel,
  \href{https://books.google.de/books?id=qIiCngEACAAJ}{Higher-Order Finite
  Element Methods}, Studies in Advanced Mathematics, Taylor \& Francis, 2003.
\newline\urlprefix\url{https://books.google.de/books?id=qIiCngEACAAJ}

\bibitem{dierckx1995curve}
P.~Dierckx, Curve and surface fitting with splines, Oxford University Press,
  1995.

\bibitem{fem_monk}
P.~Monk, P.~Peter~Monk, P.~Department~of Mathematics Sciences Peter~Monk, O.~U.
  Press, \href{https://books.google.de/books?id=zI7Y1jT9pCwC}{Finite Element
  Methods for Maxwell's Equations}, Numerical Analysis and Scientific
  Computation, Clarendon Press, 2003.
\newline\urlprefix\url{https://books.google.de/books?id=zI7Y1jT9pCwC}

\bibitem{uq_inv}
J.~Bardsley, \href{https://books.google.de/books?id=mfV1DwAAQBAJ}{Computational
  Uncertainty Quantification for Inverse Problems}, Computer Science and
  Engineering, Society for Industrial and Applied Mathematics, 2018.
\newline\urlprefix\url{https://books.google.de/books?id=mfV1DwAAQBAJ}

\bibitem{probability_theory}
A.~Klenke, Probability Theory: A Comprehensive Course, Springer London, 2013.

\bibitem{kalman}
M.~Grewal, A.~Andrews,
  \href{https://books.google.de/books?id=sZbxLK-NKb0C}{Kalman Filtering: Theory
  and Practice Using MATLAB}, Wiley, 2011.
\newline\urlprefix\url{https://books.google.de/books?id=sZbxLK-NKb0C}

\bibitem{hager1989updating}
W.~W. Hager, Updating the inverse of a matrix, SIAM review 31~(2) (1989)
  221--239.

\bibitem{RTO}
J.~Bardsley, A.~Solonen, H.~Haario, M.~Laine,
  \href{https://doi.org/10.1137/140964023}{Randomize-then-optimize: A method
  for sampling from posterior distributions in nonlinear inverse problems},
  SIAM Journal on Scientific Computing 36~(4) (2014) A1895--A1910.
\newblock \href {http://arxiv.org/abs/https://doi.org/10.1137/140964023}
  {\path{arXiv:https://doi.org/10.1137/140964023}}, \href
  {https://doi.org/10.1137/140964023} {\path{doi:10.1137/140964023}}.
\newline\urlprefix\url{https://doi.org/10.1137/140964023}

\bibitem{cde_article}
J.~{Vond{\v{r}}ejc}, H.~G. {Matthies}, {Accurate computation of conditional
  expectation for highly non-linear problems}, arXiv e-prints (2018)
  arXiv:1806.03234\href {http://arxiv.org/abs/1806.03234}
  {\path{arXiv:1806.03234}}.

\end{thebibliography}

\end{document}